\documentclass[
   ,final            
  ,numberedheadings 
  ]
  {aipproc}

\layoutstyle{8x11single}

\begin{document}

\title{CP Violation Results from CDF}


\author{Austin Napier}{
  address={Tufts University},
  address ={for the CDF Collaboration}
}



\begin{abstract}
\noindent
We present world-leading results on CP-violating asymmetries and branching fractions of several decay modes of $B^0$, $B^0_s$ , and $\Lambda_b$ hadrons into charmless two-body, and of $B^\pm$ into charm, final states collected by the CDF detector. We also report a new measurement of CP-violating asymmetries in $D^{*\pm}$-tagged $D^0 \rightarrow h^+h^-$ ($h= K$ or $\pi$) decays, where any enhancement from the Standard Model prediction would be unambiguous evidence for New Physics.

\end{abstract}

\maketitle


\noindent
CP violation is predicted by the Standard Model (SM) due to a non-zero phase angle in the CKM matrix, and it is well-established in s-quark (K-short, K-long) and b-quark (B-light, B-heavy) systems.  
CP violation is labelled ``direct" if there are different decay widths for a particle and the antiparticle and ``indirect" if it occurs as a result of ``mixing" or oscillation of one neutral meson system to another.   Hadronic decays of charm and beauty hadrons are powerful probes of flavor dynamics, and CDF has accumulated large samples of these decays.  This talk will focus on three areas: (1) charmless decays of $B^0$, $B^0_s$, and $\Lambda_b$ to two charged hadrons, (2) $B^-$ decays to $D^0 h^-$ and $\overline{D}^0 h^-$ (and charge conjugates) followed by $D \rightarrow K \pi$ ($h= K$ or $\pi$) , and (3) $D^*$-tagged decays of $D^0$ and $\overline{D}^0$ to $\pi^+ \pi^-$ and $K^+ K^-$ final states.

\section{ Two-body charmless B decays}


To study $B \rightarrow h h'$ decays, where $B = B^0$, $B^0_s$, or $\Lambda_b$ and h,h$'$ = $\pi$,K, or p, CDF uses a three-level trigger which requires two opposite-charge tracks, both with transverse momentum greater than 2 GeV/c.  We exploit the long B lifetime, vertex-pointing, hard fragmentation, and high B-mass
to select two-body B-decay candidates.  The impact parameter of the B is required to be less than 140 $\mu$m and the
transverse path length of the candidate is required to be greater than 200 $\mu$m.  This gives an overall acceptance of about 2\% for b-hadrons with $p_T > 4$ GeV/c and pseudo-rapidity ($-1 < \eta < +1$).  The trigger selection must be confirmed more accurately offline, and cuts on additional variables such as isolation and 3-D vertex quality are imposed.  The present analysis uses data from 6 fb$^{-1}$ of integrated luminosity.  Despite good mass resolution (\~~22 MeV/c$^2$) individual decay modes overlap when plotted as $M_{\pi+\pi-}$ in a single peak of width \~~35 MeV/c$^2$.  We exploit the correlation between momenta and invariant mass and particle ID information from the drift chamber (dE/dx) to determine signal composition using an extended  maximum likelihood fit.  Monte Carlo events are used to model the different signals.  The maximum likelihood fit results are shown in Fig. 1(a) and yield 10,200 $B^0 \rightarrow K^+\pi^-$, 3,008 $B^0_s \rightarrow K^+ K^-$, 2,600 $B^0 \rightarrow \pi^+ \pi^-$, 760 $B^0_s \rightarrow K^-\pi^+$, 120 $B^0 \rightarrow K^+ K^-$, and 94 $B^0_s \rightarrow \pi^+ \pi^-$.   We report the first evidence of $B^0_s \rightarrow \pi^+ \pi^-$ and measure the branching ratio:

\begin{center}
 $BR(B^0_s \rightarrow \pi^+ \pi^-) = [0.57 \pm 0.15({\rm stat}) \pm 0.10({\rm syst})] \times 10^{-6}$
\end{center}
\noindent 
at $3.7\sigma$ significance.  The significance of the $B^0 \rightarrow K^+ K^-$ decay is less than 3$\sigma$;
however, a two-sided limit of $0.05\times 10^{-6} < BR < 0.46\times 10^{-6}$ can be set at $90\%$ CL.  These two rare decays are driven by W-exchange and "penguin annihilation" diagrams, which are traditionally difficult to calculate.  
Details of the analysis are given in \cite{CDF_BS}, and previous CDF results are in \cite{CDF_preBS}.
The new results should provide a significant constraint on theoretical calculations; see \cite{Zhu}, for example.





\section{  Atwood-Dunietz-Soni (ADS) Analysis}

The ADS analysis \cite{ADS_method} provides a clean way to measure the ${\bf \gamma}$ angle in the Unitarity Triangle.  This angle is the least certain of the three angles and it is one of the last measurements necessary to insure that the CKM formalism provides the correct explanation for CP violation in the SM.  
The method uses interference between two different B$^-$ decay modes, for example, 
$B^- \rightarrow D^0 K^-$ followed by the doubly Cabibbo suppressed (DCS) decay $D^0 \rightarrow K^+ \pi^-$ and color suppressed $B^- \rightarrow \overline{D}^0 K^-$ followed by Cabibbo favored $\overline{D}^0 \rightarrow K^+ \pi^-$ both resulting in the same final state particles ($K^+ \pi^- K^-$).   The ratio of the square of the matrix elements for these two decays is proportional to the absolute value squared of $(V_{cb} V^*_{us}) / (V_{ub} V^*_{cs})$ times $BR(D^0 \rightarrow K^+\pi^-)/BR(\overline{D}^0 \rightarrow K^+ \pi^-)$.  The comparable interfering amplitudes means that large CP violation might be expected.  The ADS observables are defined as:
\begin{center}
$R_{ADS}(h) = \frac{BR(B^- \rightarrow D_{sup} h^-)+BR(B^+ \rightarrow D_{sup} h^+)}{BR(B^- \rightarrow D_{fav} h^-)+BR(B^+ \rightarrow D_{fav} h^+)}$ \hskip 6mm
$A_{ADS}(h) = \frac{BR(B^- \rightarrow D_{sup} h^-)-BR(B^+ \rightarrow D_{sup} h^+)}{BR(B^- \rightarrow D_{sup} h^-)+BR(B^+ \rightarrow D_{sup} h^+)}$
\end{center}

\noindent
where $h$ is $K$ or $\pi$,  and $D_{fav} \rightarrow K^- \pi^+$ and $D_{sup} \rightarrow K^+ \pi^-$ for $B^-$.  We report CDF results using 7 fb$^{-1}$ of integrated luminosity.  In addition to trigger requirements, cuts on isolation of the B-candidate and quality of the 3-D vertex fit are used to improve the signal to noise.  These cuts are required to see the DCS D-decays.  Results of an extended maximum likelihood fit are shown in Fig. 1(c) and 1(d) and yield:  
$N(B^- \rightarrow D_{sup}K^-) + N(B^+ \rightarrow D_{sup}K^+) = 32 \pm 12$
and 
$N(B^- \rightarrow D_{sup}\pi^-) + N(B^+ \rightarrow D_{sup}\pi^+) = 55 \pm 14$,
\noindent
providing the first evidence at $3.2\sigma$ significance of a $B^- \rightarrow D_{sup}K^-$ signal at the Tevatron.  The number of $B^- \rightarrow D_{fav}\pi^-$ decays is $\sim19,700$ and the number of $B^- \rightarrow D_{fav}K^-$ decays is $\sim1460$.
The measured ADS physics observables are:

\vskip 2mm
\begin{center}
$R_{ADS}(\pi) = [  2.8 \pm 0.7 (\rm stat) \pm 0.4 (\rm syst) ] \times 10^{-3}$ \hskip 6mm
$A_{ADS}(\pi) = [ 0.13 \pm 0.25 (\rm stat) \pm 0.02 (\rm syst) ] $
$R_{ADS}(K) = [ 22.0 \pm 8.6 (\rm stat) \pm 2.6 (\rm syst) ] \times 10^{-3}$  \hskip 6mm
$A_{ADS}(K) = [ -0.82 \pm 0.44 (\rm stat) \pm 0.09 (\rm syst) ] $
\end{center}
\vskip 2mm

\noindent
 More details are available in \cite{CDF_note_10615}.  These measurements agree well with results reported by BaBar 
 \cite{BABAR_ADS} and Belle \cite{BELLE_ADS}.






\section{CP studies of $D^0$ and $\overline{D}^0$ decays to $\pi^+\pi^-$ and $K^+K^-$}

The SM prediction for c-quark states is extremely small, however mixing is already well-established in the charm sector ($D^0$-$\overline{D}^0$); see for example \cite{CDF_mixing}.  CDF measures the time-integrated CP asymmetry:

\begin{center}
$A_{CP} = \frac {\Gamma(D^0 \rightarrow h^+h^-) - \Gamma(\overline{D}^0 \rightarrow h^+h^-)}{\Gamma(D^0 \rightarrow h^+h^-) + \Gamma(\overline{D}^0 \rightarrow h^+h^-)}$ 
\end{center}

\noindent
using $D^{*\pm}$ decays to tag $D^0$ and $\overline{D}^0$ with $h = \pi$ or $K$.   A two-track trigger is used to trigger on displaced tracks when
both $h^+$ and $h^-$ have $p_T > 2 GeV/c$ and impact parameter $> 100 \mu$m.
We use fits of the $D\pi$ mass distributions, combining the slow pion with the singly-Cabibbo-suppressed (SCS) D-decay to $h^+h^-$,   
to determine the CP asymmetries.  The distributions are weighted to account for charge and momentum dependent
detector asymmetries. 
The present results are obtained from 5.94 fb$^{-1}$ integrated luminosity, resulting in
$106,421 \pm 361$ $D^{*+} \rightarrow D^0 \pi^+ \rightarrow [\pi^-\pi^+]\pi^+$ and
$110,447 \pm 368$ $D^{*-} \rightarrow \overline{D}^0 \pi^- \rightarrow [\pi^-\pi^+]\pi^-$.
The sample of $D^0 \rightarrow K^+K^-$ decays is more than twice as large.  We find:

\vskip 2mm
\begin{center}
$A_{CP} (\pi^+ \pi^-) = [ 0.22 \pm 0.24 (\rm stat.) \pm 0.11 (\rm syst.) ] \%$ \hskip 6mm
$A_{CP} (K^+K^-) = [ -0.24 \pm 0.22 (\rm stat.) \pm 0.10 (\rm syst.) ] \%$
\end{center}
\vskip 2mm
\noindent
These results are consistent with no direct CP violation in $D^0 \rightarrow h^+h^-$ decays. More details are provided in \cite{CDF_note_10296}.

\noindent
To first order, the CP asymmetry may be written as: $A_{CP} = a^{dir} + (<t>/\tau)a^{ind}$ where $a^{dir}$ is the direct CP asymmetry and $a^{ind}$ is the indirect CP asymmetry.  $<t>$ depends on the experimental sample, sensitivities, etc., and $\tau$ is the $D^0$ lifetime.  The CDF measurements give a linear dependence with slope $-2.40 \pm 0.03$ for $\pi^+\pi^-$ and slope $-2.65 \pm 0.034$ for $K^+K^-$.  For the B-factories, with unbiased acceptance, the slope is $-1$.  A plot of $a^{dir}$ vs. $a^{ind}$ for CDF, BaBar, and Belle thus provides a constraint on the values of $a^{dir}$ and $a^{ind}$.   See Fig. 1(b).  If we assume no direct CP violation in the charm sector, then $A_{CP} = (<t>/\tau)a^{ind}$, and the measurements imply:
$a^{ind} (\pi^+ \pi^-) = [ 0.09 \pm 0.10 (\rm stat.) \pm 0.05 (\rm syst.) ] \%$  and
$a^{ind} (K^+K^-) = [ -0.09 \pm 0.08 (\rm stat.) \pm 0.04 (\rm syst.) ] \%$. 
\noindent
If these are regarded as two independent measurements of the same quantity, the value of $a_{CP}^{mix}$ =
$[ -0.01 \pm 0.06 (\rm stat.) \pm 0.05 (\rm syst.) ] \%$.

\begin{figure}
  \includegraphics[height=0.7\textheight]{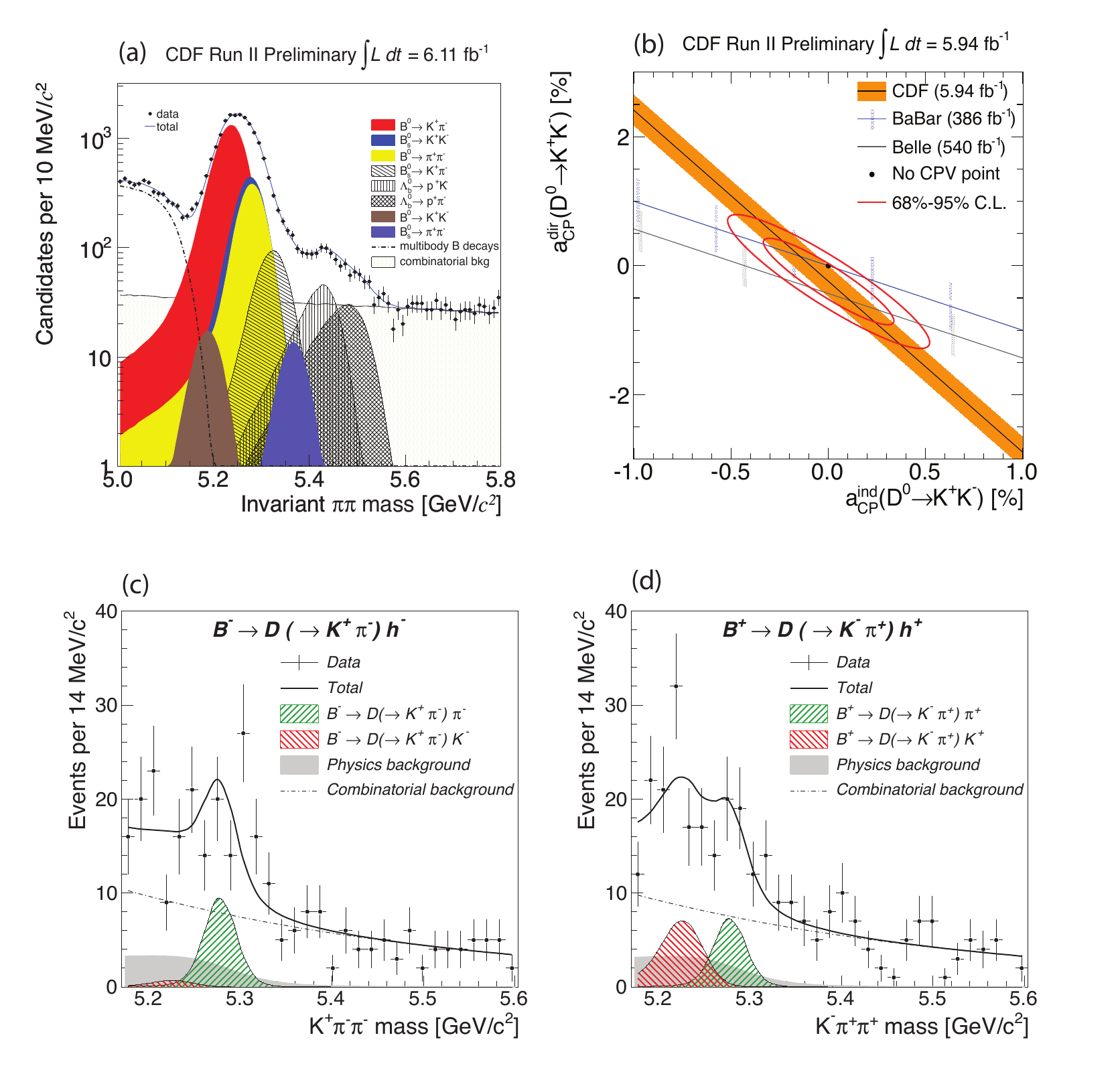}
  \caption{Invariant mass of B candidates, plotted using a $\pi$ mass for both decay products (a); variation of direct and indirect CP violation parameters in $D^0 \rightarrow K^+K^-$ and $\overline{D}^0 \rightarrow K^+K^-$ decays (b); ADS analysis results for the suppressed modes for $B^- \rightarrow Dh^-$ (c) and $B^+ \rightarrow Dh^+$ (d) with $h = \pi$ or $K$.  A $\pi$ mass is assigned to the charged track from the $B$ candidate vertex.}
\end{figure}

\section{Conclusions}

We present updated CDF results from three different analyses.   In two-body charmless $B^0$ and $B^0_s$ decays CDF observes $B^0_s \rightarrow \pi^+ \pi^-$ at $3.2\sigma$ significance and sets new limits on $B^0 \rightarrow K^+ K^-$ (``annihilation-driven" decays).  An ADS analysis of $B$ decays yields results comparable to B-factory measurements and provides a clean way to measure the angle $\gamma$ in the Unitarity Triangle.   ADS observables are measured using the decays $B^- \rightarrow D K^-$ and $B^- \rightarrow D \pi^-$ (and charge conjugate decays).  
Finally, high statistics measurements of $D^*$-tagged SCS $D^0$ and $\overline{D}^0$ decays are used to search for CP violation in the charm sector.  No evidence for direct CP violation is observed.  Updates are planned for all three analyses with the full CDF
dataset, expected to be well in excess of 10 fb$^{-1}$ integrated luminosity.

\bibliographystyle{aipproc}   

\bibliography{sample}




\end{document}